\documentclass[12pt]{article}
\usepackage{graphicx}
\usepackage{amsmath}
\usepackage{lipsum}
\usepackage{hyperref}
\usepackage{enumitem}
\usepackage{float}
\usepackage{enumitem}

\title{\textbf{Generative Voice Bursts during Phone Call}}
\author{
 Paritosh Ranjan \\
  IBM  \\
  \texttt{paranjan@in.ibm.com} \\
  \and
 Surajit Majumder \\
  IBM  \\
  \texttt{surajit.majumder@ibm.com} \\
  \and
 Prodip Roy \\
  IBM  \\
  \texttt{prodipro@in.ibm.com} \\
}
\date{\today}

\begin{document}

\maketitle

\begin{abstract}
In critical situations, conventional mobile telephony fails to convey emergency voice messages to a callee already engaged in another call. The standard “call waiting” alert does not provide the urgency or content of the waiting call. This paper proposes a novel method for transmitting Generative Voice Bursts (GVB)—short, context-aware audio messages—during ongoing calls, from either pre-authorized or dynamically prioritized callers. By leveraging generative AI techniques, the system automatically generates spoken messages from contextual inputs (e.g., location, health data, images, background noise) when the caller is unable to speak due to incapacitation or environmental constraints. The solution incorporates voice, text, and priority inference mechanisms, allowing high-priority emergency messages to bypass conventional call waiting barriers. The approach employs models such as GPT-Neo for generative text, which is synthesized into audio and delivered in configurable intervals (G seconds) and counts (N times), ensuring minimal disruption while preserving urgency. This method holds potential for significant impact across telecom, mobile device manufacturing, and emergency communication platforms.

\end{abstract}

\section{Introduction}

In conventional mobile telephony systems, when a user is already engaged in a call, an incoming call from another user is relegated to “call waiting,” with no means of content transmission unless the active call is ended. This mechanism proves inadequate in emergency scenarios where the caller needs to communicate urgent or life-critical information, particularly if the caller is physically unable to speak due to injury, shock, or ambient noise (e.g., during a natural disaster or a violent incident).

There is currently no established method or system that allows a waiting caller to send a real-time voice or synthesized message during an ongoing call. The lack of such a feature can result in delayed responses during emergencies, with potentially catastrophic consequences.

This paper introduces a novel solution that enables the generation and transmission of short voice bursts during active calls, based on either prior caller approval or dynamically inferred call priority. The core innovation lies in the use of Generative AI models (e.g., transformers like GPT-Neo) to synthesize emergency messages when the caller is incapacitated or unable to communicate effectively. The voice bursts are limited in duration (3–5 seconds) and are transmitted at intervals (G seconds), as configured by the callee, and up to a predefined number of repetitions (N times).

The system dynamically evaluates caller priority based on contextual data such as caller location, movement, time of call, and real-time physiological indicators captured via IoT devices and wearables. In high-priority situations, the system bypasses waiting status and automatically delivers voice bursts or connects the emergency call. For lower-priority messages, a text burst with a beep may be shown on the callee’s screen instead.

This approach integrates multiple modalities—speech recognition, natural language generation, context inference, and audio playback—into a cohesive emergency communication framework. It opens new possibilities for proactive response mechanisms in personal safety, healthcare, disaster response, and smart communication systems.

\section{Brief Description of the Invention}

The invention enables the transmission of short, context-aware Generative Voice Bursts (GVBs) during an ongoing phone call from either a pre-approved caller or one dynamically identified as high-priority. Using generative AI, it can create voice messages when the caller is incapacitated and unable to speak, based on contextual inputs such as location, health data, images, or ambient sound. These voice bursts, lasting 3–5 seconds, are delivered at configurable intervals and counts. The system intelligently prioritizes calls, enabling emergency communication even during call waiting, and optionally displays text bursts for lower-priority messages, enhancing responsiveness in critical situations.

Based on the determined call priority and caller’s current state: 

\begin{itemize}
    \item Generative text to be used as voice burst if the caller is incapacitated.
    \item Call gets connected for highest priority emergency.
    \item Voice Burst gets played for medium priority emergency.
    \item Text Burst with Beep (Voice converted to text and displayed on screen) for low priority emergency.
\end{itemize}

\section{Reduction to Practice}

Method to automatically generate what the person might be trying to say in case of emergency and play that as voice burst. let's take an example if the person who is in emergency is incapacitated to talk due to an emergency health situation or accident or shock or any other emergency situation, then the system can generate the emergency text based on the situation and condition of the user using Generative AI and send that text as voice note to assist the user in emergency.

During the text generation process, the AI model takes a seed input, such as images, videos, keywords etc. and uses its learned knowledge to predict the most probable words or phrases. 

Generative AI is achieved through various techniques such as:

\begin{enumerate}
    \item \textbf{Transformers:}
Transformers such as GPT-3, LaMDA, Wu-Dao, and ChatGPT mimic cognitive attention and measure the significance of input data parts. They are trained to understand language or images, learn classification tasks, and generate texts or images from massive datasets.

GPT-3 is a deep learning based language model that is trained on merely 175 billion parameters. Because of its vast amount of training parameters, it performs well on a variety of natural language processing tasks. The model is ideal for most natural language processing procedures including text generation, sentiment analysis, and dialogue models.
    \item \textbf{Generative adversarial networks (GANs):}
GANs consist of two neural networks: a generator and a discriminator that work together to find equilibrium between the two networks.
The generator network generates new data or content resembling the source data, while the discriminator network differentiates between the source and generated data to recognize what is closer to the original data.
    \item \textbf{Variational autoencoders: }
Variational auto-encoders utilize an encoder to compress the input into code, which is then used by the decoder to reproduce the initial information. This compressed representation stores the input data distribution in a much smaller dimensional representation, making it an efficient and powerful tool for generative AI.

    \item\textbf{Dataset choice for emergency generative text:}

            \begin{enumerate}
                    \item Keywords provided by the caller.
                    \item Text descriptions of the Gestures provided by the caller.
                    \item Text descriptions of the Images captured by the caller’s phone during the call.
                    \item Text descriptions of the Videos captured by the caller’s phone during the call.
                    \item Transcript of the Background speech captured by the caller’s phone during the call.
                    \item Text Description of the Background noise captured by the caller’s phone during the call.
                    \item Caller’s context i.e., time of call, location, data from IoT devices, wearables, health stats, activity stats etc.
                    \item Type of Location i.e., Highway, Hospital, Bank etc.
                \end{enumerate}

    \item\textbf{How to generate text?}
        \begin{enumerate}
            \item Install one of the natural language processing libraries — transformers.
            \item Transformers includes different natural language processing pipelines. We will be using text generation pipeline.
            \item EleutherAI / gpt-neo-2.7B: https://huggingface.co/EleutherAI/gpt-neo-2.7B
        \end{enumerate}

    \item\textbf{Transformers Installation}
    \begin{enumerate}
        \item\ pip install transformers
        \item\ from transformers import pipeline
    \end{enumerate}
    \item\textbf{Generating Your Content Creator}
        \begin{enumerate}
            \item\ gen = pipeline('text-generation', model ='EleutherAI/gpt-neo-2.7B')
        This code calls pipeline function and specifies its type i.e., text-generation, and defines the pre-trained as a keyword argument. In our case we are using GPT-Neo with 2.7 billion parameters. You may want to decrease the size depending on the speed of your computer. Below examples illustrate how to decrease parameters.
        \end{enumerate}
\item\textbf{Generating Text}

If you want to generate a text based on context, you will need to specify that context so that model can understand and generate text according to that context. Create a variable named context and write a sentence in string format.

\begin{verbatim}
context = "House Fire Help Come"
output = generator(context, max_length=50, do_sample=True, temperature=0.9)
\end{verbatim}

In the above code:
\begin{itemize}
    \item The first argument to the \texttt{generator} function is the context string.
    \item \texttt{max\_length} sets the maximum number of tokens (words/characters) in the output.
    \item \texttt{do\_sample=True} enables sampling to allow for more diverse outputs.
    \item \texttt{temperature=0.9} controls the randomness of the output (higher values yield more randomness).
\end{itemize}

\textbf{Generated Text Example:}
\begin{quote}
\emph{``The house is on fire. Please call the fire brigade and ambulance as soon as possible. Please come back.''}
\end{quote}

Since the output is returned in dictionary format, it can be written to a text file for further use or editing:

\begin{verbatim}
with open('dl.txt', 'w') as f:
    f.write(str(output))
\end{verbatim}

\subsection*{Determining Caller Incapacity}

Caller incapacity is determined through multiple AI/ML-based modalities, including:

\begin{enumerate}
    \item Analysis of images or videos captured by the caller's device using computer vision techniques.
    \item Analysis of textual descriptions or transcripts derived from the captured media.
    \item Gesture-based inputs provided by the caller.
    \item Detection of specific keywords from the caller such as \texttt{Help} or \texttt{Can't Speak}.
    \item Recognition of silence or absence of speech during a permitted voice burst in a suspected emergency situation.
\end{enumerate}

\begin{figure}
    \centering
    \includegraphics[width=0.75\linewidth]{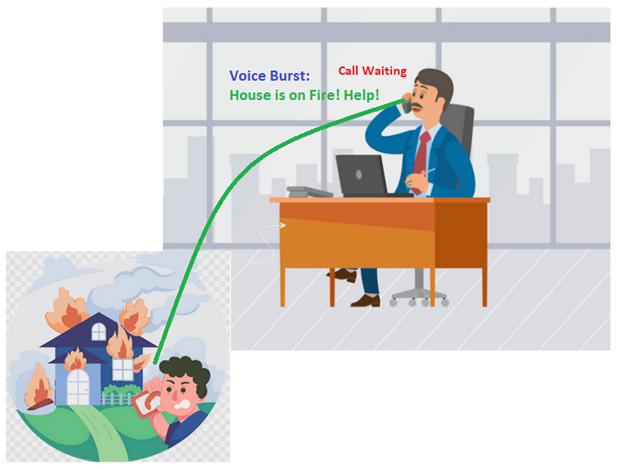}
    \caption{Voice-Burst-from-a-pre-approved-call-waiting-user}
    \label{fig:Voice-Burst-from-a-pre-approved-call-waiting-user}
\end{figure}

\item\textbf{Method to Capture and Play Voice Burst During a Call for a Pre-approved Call Waiting Caller}

Consider a scenario where phone user A is the father of phone user C, and user C is a child alone at home. The method below outlines the process of capturing and playing a voice burst from a pre-approved call waiting caller during an ongoing call:

\begin{enumerate}
    \item Phone user A is already engaged in a call with phone user B.
    \item Phone user C encounters an emergency situation and attempts to call phone user A.
    \item Phone user C’s call is placed into the call waiting queue on phone user A’s device.
    \item Phone user A has previously pre-approved phone user C to send voice bursts during call waiting.
    \item While in the waiting state, phone user C is permitted to transmit short voice bursts of 3 to 5 (or $t$) seconds to phone user A.
    \item Phone user C sends a voice burst describing the emergency situation. For example, the message may include:
    \begin{itemize}
        \item ``The house is on fire.''
        \item ``I am fainting.''
        \item ``A thief has entered the house.''
        \item ``I have met an accident.''
    \end{itemize}
    \item The voice burst provides enough contextual urgency for phone user A to assess the criticality of the situation.
    \item Based on this assessment, phone user A may decide to end the ongoing call with phone user B and respond immediately to user C.
    \item After a configurable gap of $G$ seconds (as set by phone user A), user C is permitted to send another voice burst if necessary.
    \item The number of allowed voice bursts $N$ is predefined and configured by phone user A.
\end{enumerate}

\item\textbf{Method to automatically allow voice burst at runtime for higher priority call with call priority detected at runtime}

Phone users may either explicitly pre-approve specific contacts to send voice bursts of 3 to 5 seconds during call waiting, or the system may dynamically determine whether to allow such voice bursts at runtime based on contextual priority analysis.

Dynamic permission for voice burst transmission is determined using the following contextual factors:

\begin{itemize}
    \item \textbf{Location:} The caller typically makes calls from home, but the current call is being made from an unusual or high-risk location such as a hospital, highway, or isolated area.
    
    \item \textbf{Timing:} The caller initiates a call at an unusual time, deviating significantly from their historical calling pattern (e.g., during late-night hours).
    
    \item \textbf{Health:} Health metrics, such as elevated heart rate detected via a wearable device (e.g., smartwatch), indicate a potential emergency situation compared to previous baselines.
    
    \item \textbf{Activity:} The caller is actively moving (as inferred from the device's motion sensors or GPS), which is atypical for their calling behavior, possibly suggesting distress.
\end{itemize}

The system assesses the caller’s context using multimodal data sources, including time-of-call, geolocation, device sensor data, and physiological inputs from IoT-enabled wearables. If the computed emergency score exceeds a predefined threshold, the system automatically overrides the default call waiting behavior. In such cases, the call is promoted in priority, the voice burst is allowed, or the emergency call is connected directly, ensuring timely response by the callee.

\begin{figure}
    \centering
    \includegraphics[width=0.9\linewidth]{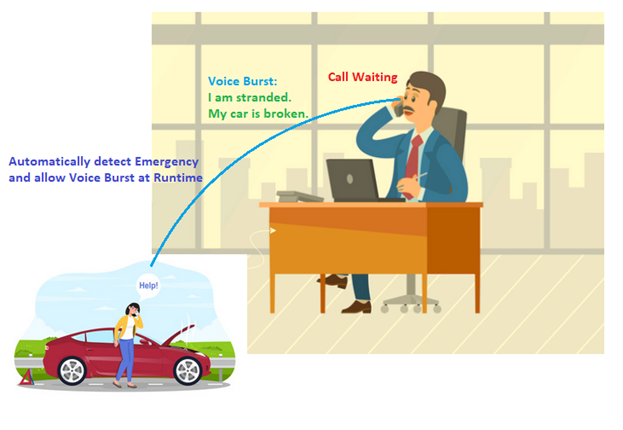}
    \caption{Allowing-voice-burst-at-runtime-after-detecting-emergency-situation}
    \label{fig: Allowing-voice-burst-at-runtime-after-detecting-emergency-situation}
\end{figure}

\end{enumerate}

\section{Advantages of the Invention}

The proposed system for enabling \textbf{Generative Voice Bursts (GVBs)} during ongoing phone calls offers several significant advantages over conventional telecommunication protocols and emergency handling methods:

\begin{enumerate}
    \item \textbf{Enhanced Emergency Communication During Active Calls:}  
    Allows critical voice information to be transmitted while the callee is on another call, overcoming the limitations of standard call waiting notifications.

    \item \textbf{Automatic Emergency Detection Using Multimodal Data:}  
    Utilizes location data, wearable sensor inputs, audio context, and historical behavior patterns to assess the urgency of the call dynamically.

    \item \textbf{Support for Incapacitated Callers via Generative AI:}  
    Employs transformer-based models to generate emergency voice messages when the caller is unable to speak due to injury, shock, or environmental constraints.

    \item \textbf{Configurable and Controlled Voice Burst Mechanism:}  
    Enables voice bursts of 3--5 seconds, delivered at user-defined intervals ($G$ seconds) and limited to a maximum number of attempts ($N$), reducing disruption.

    \item \textbf{Non-Invasive Alerts for Low-Priority Events:}  
    In less urgent cases, the system converts the message to text and displays it on the call screen with an audible beep, avoiding interruption of the ongoing call.

    \item \textbf{Customizable Permissions and Privacy Control:}  
    Provides users the option to pre-authorize trusted contacts for voice bursts, ensuring security and personal control over communication permissions.

    \item \textbf{Applicability Across Multiple Domains:}  
    The invention is versatile and can be integrated into telecom infrastructure, mobile devices, calling applications, and emergency management platforms.

    \item \textbf{Reduced Response Time in Critical Situations:}  
    Facilitates faster awareness and reaction from the receiver, which can be vital in life-threatening scenarios or time-sensitive emergencies.
\end{enumerate}

\section{Conclusion}
This paper presents a novel system that enables the generation and transmission of Generative Voice Bursts (GVBs) during ongoing phone calls, addressing a critical gap in real-time emergency communication. By leveraging generative AI, multimodal context analysis, and controlled voice burst injection, the system empowers callers—especially in emergency or incapacitated states—to communicate essential information without requiring the termination of active calls.

Unlike traditional call waiting, this approach introduces intelligent call prioritization, allowing high-urgency communications to bypass waiting queues based on real-time assessments of the caller’s physical, environmental, and contextual state. The system is configurable, privacy-respecting, and broadly applicable across mobile, telecom, and VoIP ecosystems.

Incorporating this technology has the potential to significantly reduce response times in critical situations, improve emergency outcomes, and enhance the responsiveness and safety features of modern communication systems. Future work may include real-world deployment trials, optimization of model inference latency, and tighter integration with emergency services and smart devices.

\section{Acknowledgment}

We would like to express our sincere gratitude to all individuals and organizations who have contributed to the success of this research. We acknowledge the invaluable support from the IBM team, whose resources and expertise have greatly enhanced this project.
Special thanks to Prodip Roy (Program Manager IBM) for their insightful feedback, guidance, and encouragement throughout the development of this work.

\section{References}
\renewcommand\refname{}

\end{document}